\documentclass[%
aip,
jcp,%
reprint,%
 groupedaddress
]{revtex4-1}

\usepackage{graphicx}
\usepackage{dcolumn}
\usepackage{bm}
\usepackage{amsmath}
\usepackage{setspace}
\usepackage{color}
\usepackage{float}
\usepackage{subfigure}
\usepackage{array}
\usepackage{verbatim}

\begin{document}


\title[]{Dynamic Heterogeneity in Crossover Spin Facilitated Model of Supercooled Liquid and Fractional Stokes-Einstein Relation}

\author{Seo-Woo Choi}
\thanks{These authors contributed equally to this work.}
\affiliation{Department of Chemistry, Seoul National University, Seoul 151-747, Korea}

\author{Soree Kim}
\thanks{These authors contributed equally to this work.}
\affiliation{Department of Chemistry, Seoul National University, Seoul 151-747, Korea}

\author{YounJoon Jung}
\email{yjjung@snu.ac.kr}
\affiliation{Department of Chemistry, Seoul National University, Seoul 151-747, Korea}

\date{\today}

\begin{abstract}
Kinetically constrained models (KCMs) have gained much interest as models that assign the origins of interesting dynamic properties of supercooled liquids to dynamical facilitation mechanisms that have been revealed in many expreiments and numerical simulations.
In this work, we investigate the dynamic heterogeneity in the fragile-to-strong liquid via Monte Carlo method
using the model that linearly interpolates between the strong-liquid like behavior and the fragile-liquid like behavior by an asymmetry parameter $b$.
When the asymmetry parameter is sufficiently small, smooth fragile-to-strong transition is observed both in the relaxation time and the diffusion constant.
Using these physical quantities, we investigate fractional Stokes-Einstein relations observed in this model.
When $b$ is fixed, the system shows constant power law exponent under the temperature change,
and the exponent has the value between that of the FA model and the East model.
Furthermore, we investigate the dynamic length scale of our systems and also find the crossover relation between the relaxation time.
We ascribe the competition between energetically favored symmetric relaxation mechanism and entropically favored asymmetric relaxation mechanism to the fragile-to-strong crossover behavior.
\end{abstract}

\maketitle

\section{Introduction}
Supercooled liquid is a metastable state of matter that appears below the melting temperature of liquid.\cite{angell1995formation,ediger1996supercooled,ritort2003glassy,debenedetti2001supercooled,cavagna2009supercooled}
One of the intriguing features of the supercooled liquid is rapidly growing relaxation times at low temperatures.
Using the temperature dependent behavior of the relaxation time, they can be classified into two different cases.
Both the viscosity and relaxation times of the strong liquids show Arrhenius behavior, while those of the fragile liquids show super-Arrhenius behavior.\cite{angell1995formation,debenedetti2001supercooled,cavagna2009supercooled}
The Arrhenius behavior shows that the effective energy barrier is constant under the temperature variation.
In the fragile liquids, however, the effective energy barrier increases as the temperature decreases. 
Many theoretical and numerical studies indicate that the energy barrier of fragile liquids is inversely proportional to the temperature, ${\Delta}E\sim A/T$,\cite{sollich1999glassy,buhot2002crossover,keys2011excitations}
which yields the log of the relaxation time that is inversely proportional to the square of temperature, ${\tau}\sim e^{A/T^2}$.

Various models have been proposed to explain and to analyze the properties of the supercooled liquids.
Among these models, kinetically constrained models (KCMs) are particularly important and have attracted 
much scientific interest because it can explain the dynamic properties of supercooled liquids despite its simplicity.\cite{buhot2001crossover,buhot2002crossover,pan2005decoupling,jung2004excitation,jung2005dynamical,pan2005decoupling,garrahan2003coarse}
KCMs are lattice models which are designed based on the dynamical facilitation mechanisms, where the thermodynamic interactions are suppressed.
Each lattice site represents spatially coarse-grained liquids, which is assumed to have length scale larger than the interaction range of molecules.
Only with simple kinetic constraints in the absence of complicated static interactions,
KCMs have been recognized as useful theoretical models that are able to describe the distinctive properties of the supercooled liquids.

The fragile-to-strong crossover behavior has been experimentally observed in confined waters and many glass forming liquids.
\cite{ito1999thermodynamic,farone2004fragile,chen2006experimental,zhang2010fragile,mallamace2010transport}
In many of these studies, it has been considered that the local structure change causes the fragile-to-strong crossover.
To model the crossover behavior, many different theoretical crossover models have been proposed using KCMs.
\cite{buhot2001crossover,buhot2002crossover,garrahan2003coarse,pan2005decoupling}
Among the models, Buhot et al. used a combination rule of symmetric and asymmetric facilitations to realize crossover behavior based on the Frederickson-Andersen (FA) model and the East model.\cite{buhot2001crossover,buhot2002crossover} 
In their study, the crossover temperature depends on the asymmetric parameter which controls the extent of asymmetricity.

In this study, we use a similar model system as used in Ref.\citenum{buhot2001crossover},  
and analyze the crossover behavior of the relaxation time, the diffusion constant and the dynamic length scale. 
We also study the heterogeneous dynamics found in the crossover models which is an important dynamic property of supercooled liquid systems. 
Investigating the relation between the relaxation time and the diffusion constant, we find the fractional Stokes-Einstein relation in this model.

The contents in this article are organized as follows: 
In Section II, we introduce the crossover model system.
In Section III, various physical quantities showing crossover behavior are calculated.
The breakdown of Stokes-Einstein relation and the power behavior of the dynamic length scale between the relaxation time are also demonstrated.
Finally, concluding remarks follow in Section IV.

\section{Theory and Computational Method}
Theories based on kinetically constrained models grant a privileged role to dynamics in explaining most properties of supercooled liquids, without invoking explanations based on structural properties.\cite{berthier2011dynamical}
One family of the models in this field is spin facilitated model, first proposed by Frederickson and Andersen.
\cite{fredrickson1984kinetic}
In this model, the liquid is described as a lattice of spins that can either take value of 1 or 0. 
We choose 1 for the active region and 0 for the inactive region. 
Spins that satisfy a certain constraint in the neighboring configuration can undergo flipping transition under the condition of detailed balance. 
The simplest version requires only one adjacent active spin. 
This model, called FA model, has inspired many variations of spin facilitated models. 
One variation, called the East model,\cite{jackle1991hierarchically} incorporates directional facilitation; a spin can flip only if there is an adjacent active spin to a pre-specified direction, say, to the East direction. 
The FA model is a well known model for describing strong liquids and the East model shows the properties of fragile liquids in terms of the temperature dependence of the relaxation times.\cite{jung2004excitation,jung2005dynamical}

We use an one-dimensional crossover model that interpolates between the FA-like model and the East model according to the asymmetry parameter, $0\le b \le1/2$, which was proposed by Buhot et al.
\cite{buhot2001crossover}
The equilibrium Hamiltonian of the system is trivial; if $n_i$ stands for the $i$-th spin, $H=\sum_{i=0}^{N-1} n_i \hspace{0.1cm}(n_i=1,0)$. The equilibrium concentration $c$ of active spins is
\begin{equation}
c = \frac{e^{-1/T}}{1+e^{-1/T}}.
\end{equation}
For convenience, the Boltzmann constant is taken to be unity. In our calculation, the system size is set to contain 100 active spins. 
The probability $P_i$ for the $i$-th spin to flip is given by
\begin{equation}
P_i = (bn_{i+1}+(1-b)n_{i-1})\text{min}\left\{1,\text{exp}\left(-\frac{{\Delta}H}{T}\right)\right\}.
\label{condition_eq}
\end{equation}
Note that the $b$ = 0 case coincides with the East model, while the $b$ = 1/2 case is symmetric like the genuine FA model. 
There is a subtle difference between the $b$ = 1/2 case and the FA model, which differs by factor of 2 in the relaxation time in the low temperature limit.
Therefore, from now on, we will use labels FA and East in place of $b$ = 1/2 and $b$ = 0, respectively. 
To calculate temporal evolution of the system, a simple version of continuous-time Monte Carlo method, called the $n$-fold way method is used.
\cite{bortz1975new}

\section{Results and Discussion}

\subsection{Dynamic heterogeneity and the breakdown of Stokes-Einstein relation}
An important dynamic property of supercooled liquid is dynamic heterogeneity.
\cite{garrahan2002geometrical,hedges2007decoupling,darst2010dynamical,berthier2011dynamical}
In supercooled liquids, particles with similar mobilities cluster together, giving rise to a mosaic of microscopic regions with different dynamics.
This spatio-temporal correlation is dubbed dynamic heterogeneity.
Common manifestations of dynamic heterogeneity are stretched exponential decay of correlation functions and the decoupling of the diffusion constant and structural relaxation.
\cite{lacevic2003spatially,kumar2006nature}
At low temperatures, local regions where structural relaxations are fast make dominant contribution to the diffusion constant, while local regions where structural relaxations are slow and have long relaxation time determine the relaxation times of the whole system. 
As a result, the Stokes-Einstein relation, $D{\tau}\sim \text{const}$, satisfied in normal liquids, does not hold in the supercooled regime.

Dynamic heterogeneities of either strong or fragile liquids have been studied thoroughly using computational models.
In contrast,
dynamic heterogeneities of systems showing the fragile-to-strong transition have been investigated to much less degree in terms of the variety in models and observables.\cite{buhot2001crossover,buhot2002crossover,pan2005decoupling} 
In this work, we investigate physical properties closely connected to dynamic heterogeneities of the fragile-to-strong crossover model. 
In the following paragraph, results of the relaxation time and the diffusion constant will be shown. 
Then, the fractional Stokes-Einstein relation and corresponding power law exponent will be presented.

First, we use the mean persistence time, ${\tau}_{\text{pers}}$, as the relaxation time of the system. 
The persistence time of the $i$-th spin is defined as the time interval between the beginning of the simulation and the first flip of that spin.
\cite{jung2004excitation} 
The mean persistence time is calculated by averaging over the whole systems and 10,000 independent trajectories.

\begin{figure*}[t]
\begin{center}
\mbox{
\hspace{-10pt}
\subfigure{
\includegraphics[angle=0,width=8.5cm]{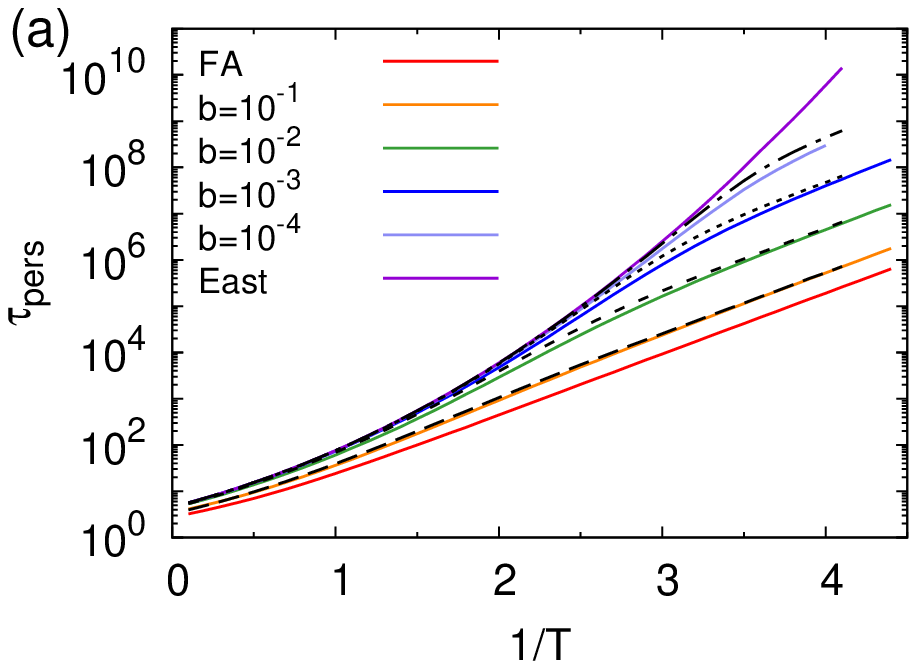}
}
\hspace{-20pt}
\subfigure{
\includegraphics[angle=0,width=8.5cm]{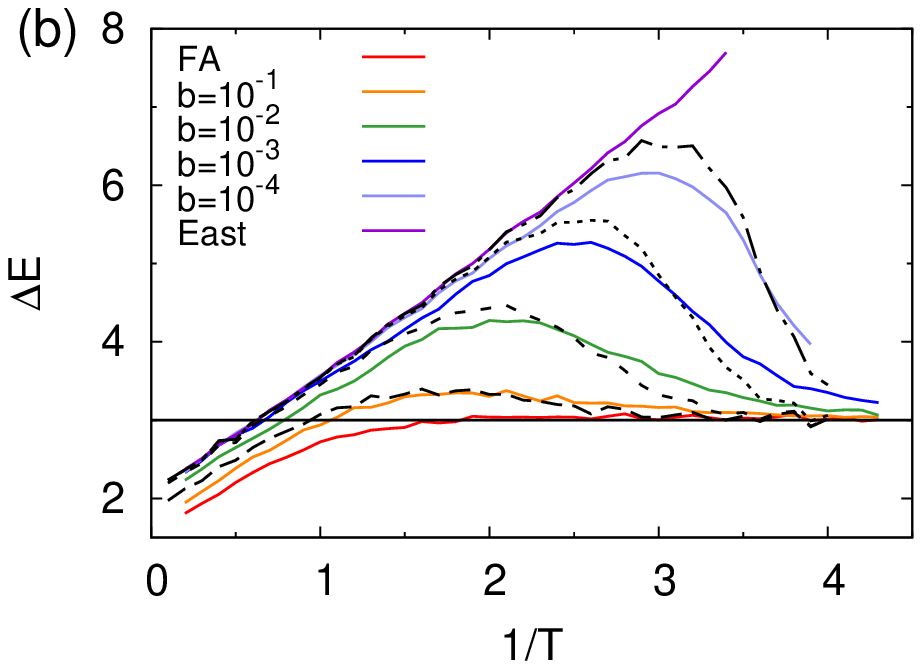}
}
}
\end{center}
 \vspace{-10pt}
 \caption{\label{fig1}
 (a) The mean persistence times of the models with various values of the asymmetric parameter $b$. 
 The black curves are the values calculated using Eq.(4). 
 The crossover behavior is observed in the models with $b=10^{-1},10^{-2},10^{-3},10^{-4}$. 
 (b) Effective energy barrier of the relaxation process. For the $b$ = 0 model, the energy barrier is constant at sufficiently low temperatures, $\tau \sim c^{-3}\sim e^{3/T}$. 
 Energy barrier of the East model is inversely proportional to the temperature. 
 Other crossover models show a peak which could be used for defining crossover temperature, $T_c$.
 Note that the derivatives of ${\tau}_{\text{pers}}$ obtained using Eq.(4) well describe each system.
 }
 \end{figure*}

${\tau}_{\text{pers}}$ of the crossover model is plotted against temperature in Fig.1(a).
The $b$ = 0 model reproduces fragile behavior and the behavior of $b$ = 1/2 model is same as that of the true FA model, which is ${\tau}_{\text{pers}}\sim c^{-3}$.
\cite{jung2004excitation}
Intermediate models with $b$ between 0 and 1/2 show super-Arrhenius behavior in high temperature regime, 
but below a certain crossover temperature, $T_c$, their relaxation times follow the Arrhenius law. 
Fig.1(b) shows this in clear fashion by plotting effective energy barrier of relaxation,
\cite{buhot2001crossover}
\begin{equation}
{\Delta}E = \frac{\text{d}\text{ln}{\tau}_{\text{pers}}}{\text{d}(1/T)}.
\end{equation}
These are in accordance with the results obtained by Buhot et al.,\cite{buhot2001crossover} in spite of different definition of the relaxation time.

If the relaxation mechanism of the intermediate model is additive with respect to the symmetric and asymmetric processes,
then the relaxation time of the intermediate models can be expressed in the following way.
\begin{equation}
\frac{1}{{\tau}_{\text{pers}}}\approx \frac{4b(1-b)}{{\tau}_{\text{sym}}}+\frac{1-2b}{{\tau}_{\text{asym}}},\hspace{0.5cm}\left(0\le b\le 1/2\right)
\label{per_eq}
\end{equation}
where ${\tau}_{\text{sym}}$ is the relaxation time of the symmetric model (FA model) and ${\tau}_{\text{asym}}$ is the relaxation time of the asymmetric model (East model).
The derivation of this equation is sketched in the appendix.

A similar equation has been proposed and used in Ref.\citenum{buhot2001crossover}.
The overall agreement between Eq.(4) and simulation results is excellent as shown in Fig.1.

\begin{figure}[t]
\includegraphics[angle=0,width=8.5cm]{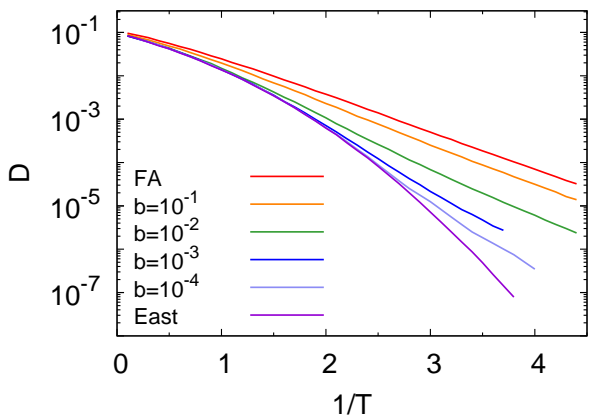}

 \vspace{-10pt}
 \caption{\label{fig2}
The diffusion constant also shows fragile-to-strong crossover behavior. However, the crossover behavior is much slower compared to the mean persistence time (Fig.1(a)).
 }
\end{figure}

\begin{figure}[t]
\includegraphics[angle=0,width=8.5cm]{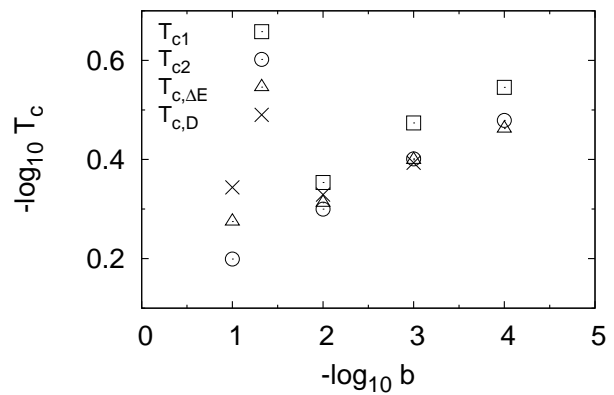}

 \vspace{-10pt}
 \caption{\label{fig3}
The crossover temperatures with different definitions. $T_{c1}$ is obtained using the definition in Ref.\citenum{buhot2001crossover}.
$T_{c2}$ is the temperature of inflection point in the Eq.(4). 
$T_{c,{\Delta}E}$ is the position of the peak in the Fig.1(b). 
$T_{c,D}$ is defined in the same way as $T_{c,{\Delta}E}$ using data of Fig.2.
 }
 \end{figure}

\begin{figure*}[t]
\mbox{
\hspace{-40pt}
\subfigure{
\includegraphics[angle=0,height=5.3cm]{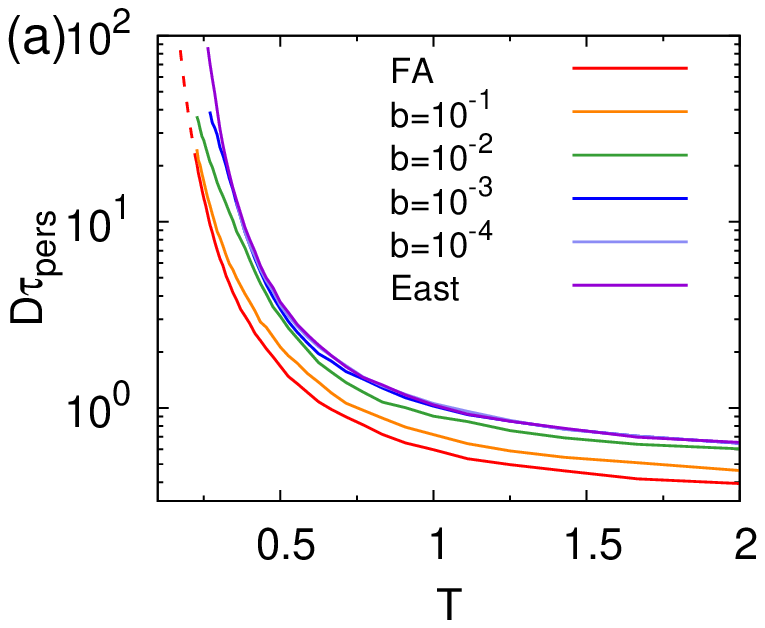}
}
\hspace{-57pt}
\subfigure{
\includegraphics[angle=0,height=5.3cm]{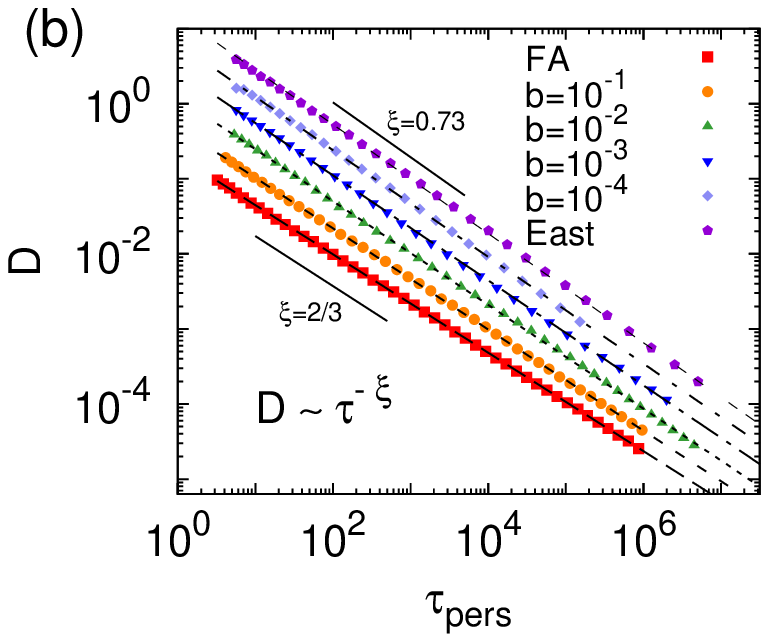}
}
\hspace{-54pt}
\subfigure{
\includegraphics[angle=0,height=5.3cm]{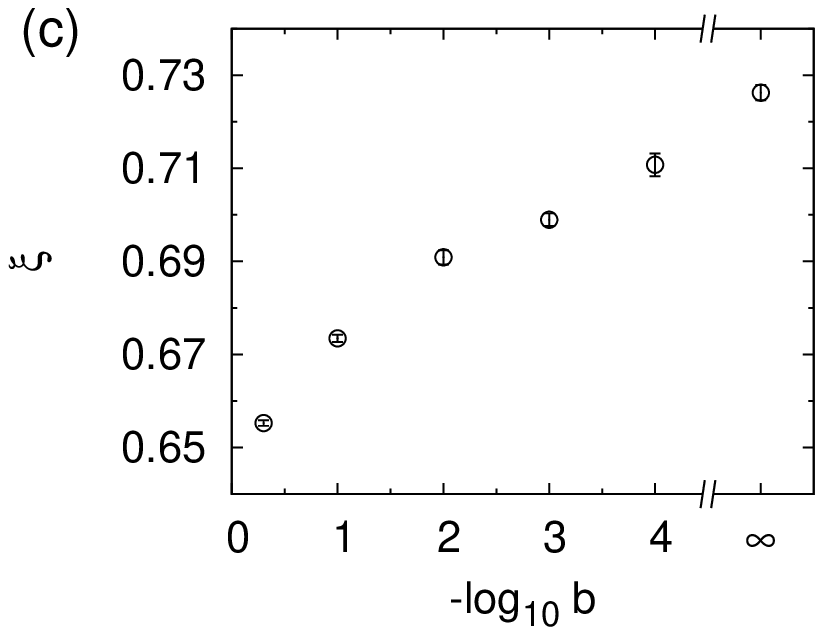}
}
}
\mbox{
\subfigure{
\includegraphics[angle=0,height=5.3cm]{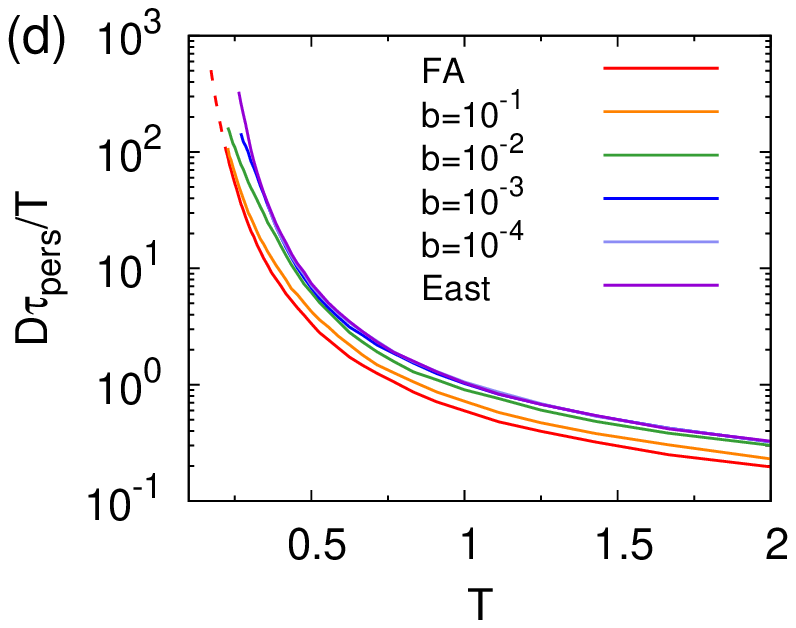}
}
\subfigure{
\includegraphics[angle=0,height=5.3cm]{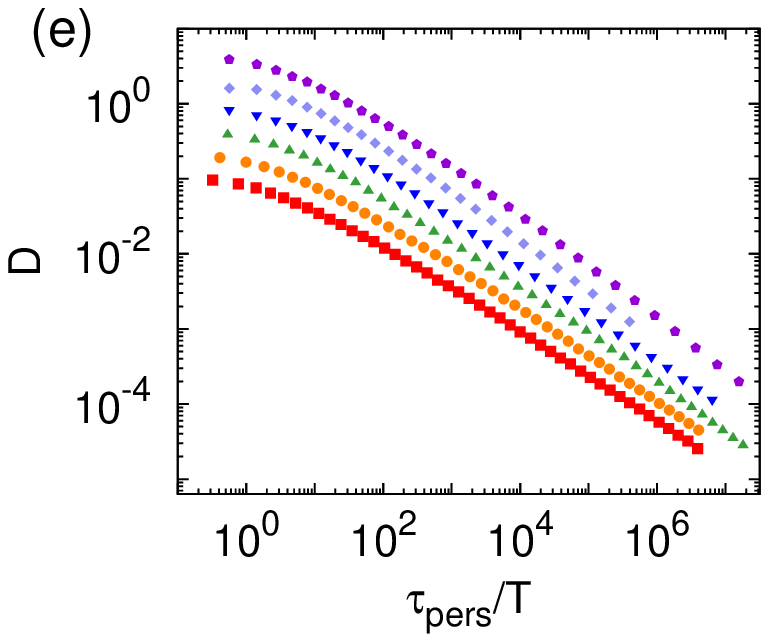}
}
}
 \vspace{-10pt}
 \caption{\label{fig4}
(a) Breakdown of the Stokes-Einstein relation and fragile-to-strong dynamic crossover. 
(b) The fractional Stokes-Einstein relation, $D\sim \tau^{-\xi}$. 
The data other than those belonging to the FA($b$ = 1/2) model are offset by factors of $10^{1/3}$ for clear comparison. 
(c) The increase of the fractional Stokes-Einstein exponent by the increase in asymmetry. 
The East model's value is also marked on the right end. 
The standard errors are estimated from simple linear regressions.
(d) Temperature dependence of $D{\tau}_{\text{pers}}/T$. As the temperature decreases, $D{\tau}_{\text{pers}}/T$ increases abruptly.
(e) Log-log plot of $D$ versus ${\tau}_{\text{pers}}/T$. At high temperatures, power law fitting is not perfect. However, as the temperature decreases, the power law is recovered. 
 }
\end{figure*}

Second, the diffusion constant is calculated by inserting the probe particles in the model and tracking the displacements of the probe particles.
\cite{jung2004excitation}
The probe does not interact with the background liquids or other probes. 
The probe makes an attempt to move to neighbor lattice sites in every 1 MCS. 
The diffusion constant is defined as in the following equation,
\begin{equation}
D=\lim_{t\rightarrow\infty}\frac{\langle[\Delta \mathbf{r}(t)]^2\rangle}{2t},
\end{equation}
where $\Delta \mathbf{r}(t)=\mathbf{r}(t)-\mathbf{r}(0)$ and the bracket is the average of the term inside over different probes and trajectories. 
The square of the displacement is averaged about 10 probes and 2,400 independent trajectories. 
Simulation time of $10^4\sim10^8$ MCS is sufficient for our temperature range. 
The dynamic crossover of the diffusion constant that is similar to that of the relaxation time could be observed (Fig.2).
However, the convergence to the FA-like behavior
\cite{jung2004excitation}
of $D\sim c^{-2}$ is slower than the case of the mean persistence time and the simple analytical model like Eq.(4) is not found for the diffusion constant.
Note that this crossover in the diffusion constant also happens in triangular lattice gas model.\cite{pan2005decoupling}

To account for the crossover temperature, different definitions are formulated and compared with each other. 
One possible definition utilized in Ref.\citenum{buhot2001crossover}
is the temperature at which the rate of the symmetric process and asymmetric processes are equal, $T_{\text{c1}}$.
Another possible choice is the temperature at the inflection point, and such temperatures on our analytic curve based on Eq.(4), $T_{\text{c2}}$, and numerically obtained curve on $\tau_{\text{pers}}$ and diffusion constant are denoted respectively as $T_{\text{c},{\Delta}E}$ and $T_{\text{c},D}$. 
As the asymmetric parameter decreases, the crossover becomes more apparent and different definitions tend to agree with each other, with the exception of the $T_{\text{c1}}$ (Fig.3).

Using the mean persistence time and the diffusion constant, the breakdown of Stokes-Einstein relation is investigated as shown in Fig.4(a).
At high temperatures, $D{\tau}_{\text{pers}}$ is nearly constant. 
However, as the temperature decreases, $D{\tau}_{\text{pers}}$ increases abruptly in all models. 
Furthermore, the fragile-to-strong dynamic crossover is visible. 
This result is consistent with that obtained from triangular lattice gas.
\cite{pan2005decoupling}

When the diffusion constant is plotted against the relaxation time (Fig.4(b)), they seem to be related by fractional exponent, $D\sim{\tau}_{\text{pers}}^{-{\xi}}$, for all models.
Note that the log-log plot does not show a clear crossover behavior in the crossover models. 
The power law exponent, ${\xi}$, does not change under the range, while it has the value between that of the FA model ($\sim$0.67) and the East model ($\sim$0.73).
\cite{jung2004excitation}
We expect that simulations in higher dimension where the difference between the exponent of the FA model and the exponent of the East model is larger
\cite{buhot2001crossover}
 would reveal the transition appearing at a single crossover temperature.
Note that, in higher dimensions, the FA model which has a finite upper critical dimenion recovers Stokes-Einstein relation,\cite{jack2006mappings,whitelam2005renormalization}
while the East model is expected to exhibit the exponent lower than 1.\cite{jung2004excitation,berthier2005length}
In Fig.4(c), the average exponents are shown. 
The exponents smoothly change from the value of the FA model to the value of the East model.

In some cases, the condition that $D{\tau}/T$ is constant over temperature change is used for the Stokes-Einstein relation.\cite{kumar2007relation,xu2009appearance}
We also show the temperature dependence of $D{\tau}_{\text{pers}}/T$ in Fig.\ref{fig4}(d), which shows similar result as Fig.\ref{fig4}(a).
When $D$ is scaled with ${\tau}_{\text{pers}}/T$, the power law relation does not hold as strongly as in the $D{\tau}_{\text{pers}}$ case at high temperatures as shown in Fig.\ref{fig4}(e).
As the temperature decreases, however, $D$ and ${\tau}_{\text{pers}}/T$ show power law behavior similar to Fig.\ref{fig4}(b).
In the asymptotic limit that the temperature is low enough, the power exponent would be eventually the same as one obtained in Fig.\ref{fig4}(b), because $\text{log}{\tau}_{\text{pers}}$ increases much faster than $\text{log}T$.

\begin{figure}[t]
\begin{center}
\subfigure{
\includegraphics[angle=0,width=8.5cm]{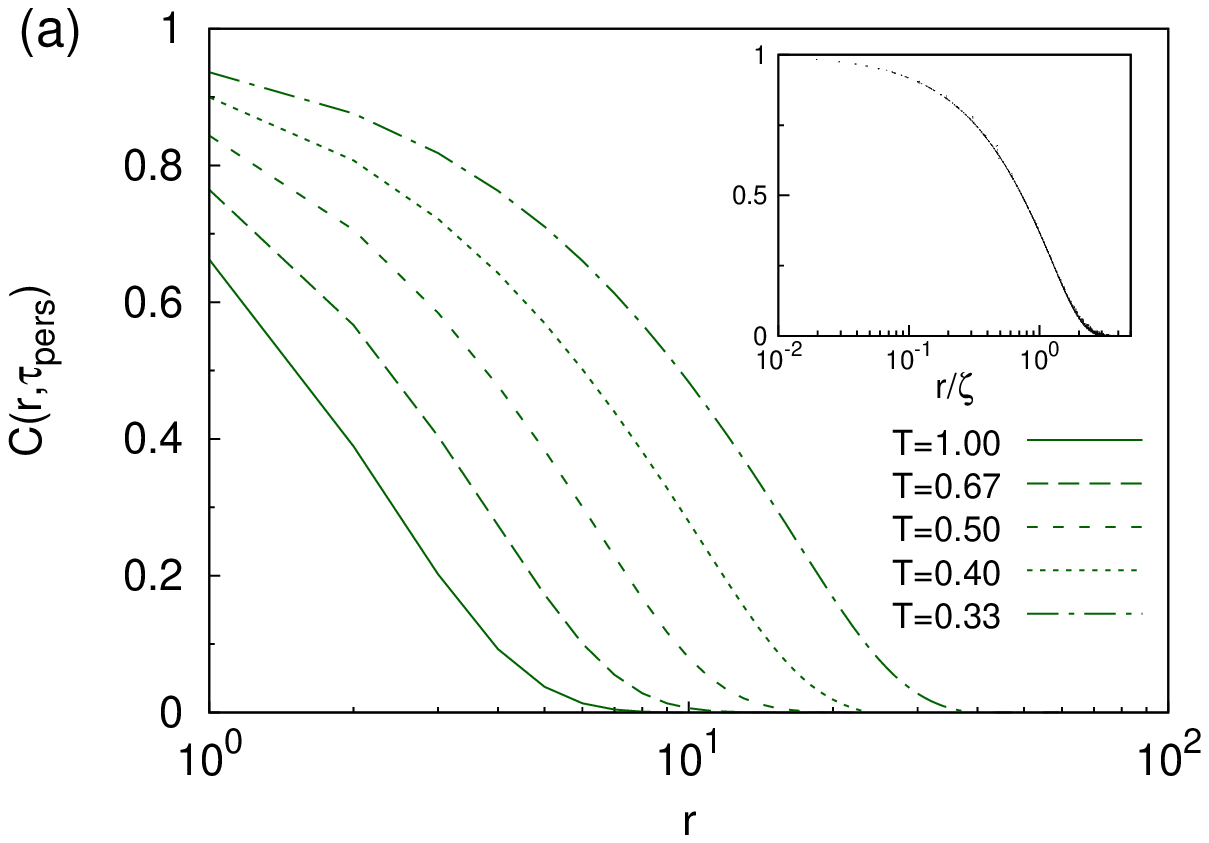}
}
\mbox{
\hspace{1pt}
\subfigure{
\includegraphics[angle=0,width=8.3cm]{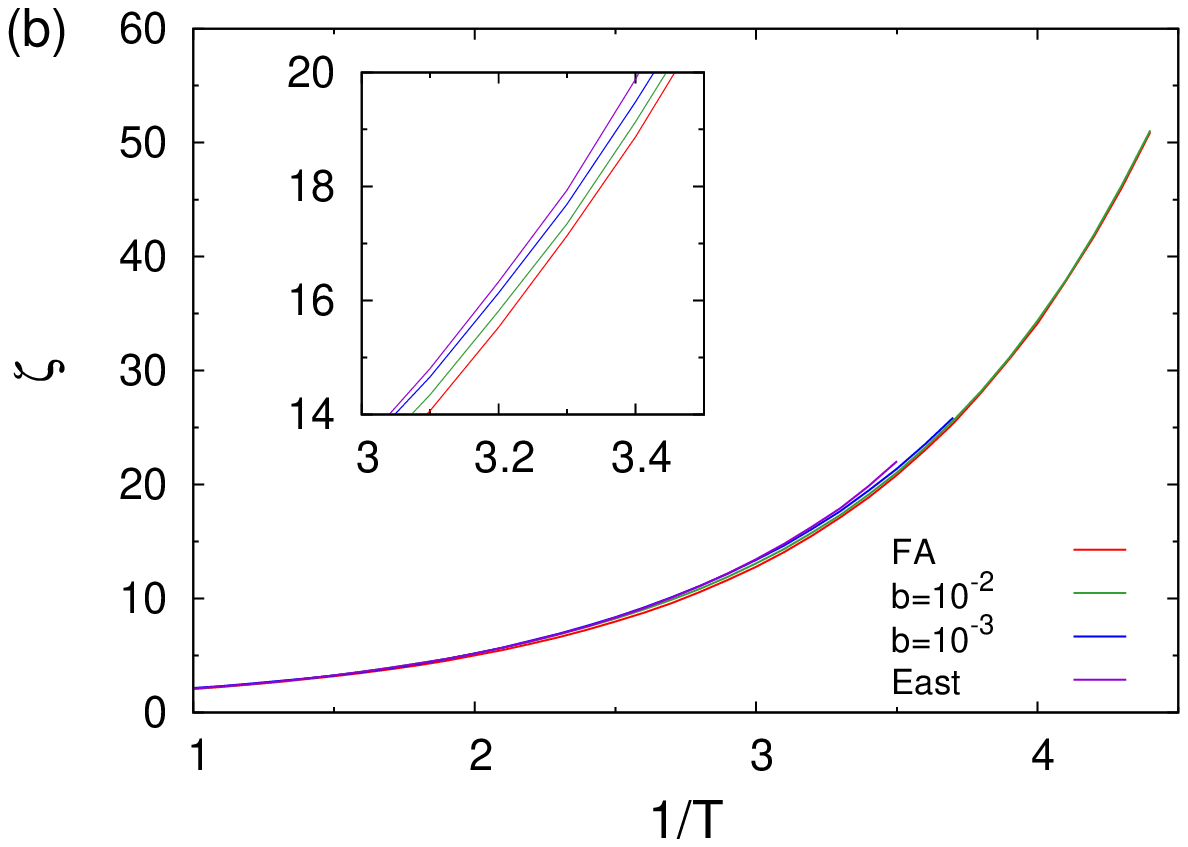}
}
}
\end{center}

 \vspace{-10pt}
 \caption{\label{fig5}
(a) Correlation function calculated with parameter $b$=0.01. Inset shows the collapse of different sets of parameters into a single curve.
(b) Dynamic correlation length is increasing as the temperature decreases. Inset is the magnification of data to see clear crossover behavior.
}

\end{figure}

\subsection{Dynamic correlation lengths}

The mobility of the spin is estimated by the persistence function ${\pi}_{i}(t)$ that is 1 if spin $i$ has never flipped, or 0 if spin $i$ has flipped at least once. 
The four-point correlation function $C(r,t)$ defined from the persistence function contains the information on the length and time scale of the dynamic heterogeneity:
\begin{equation}
C(r,t) = \frac{\langle \frac{1}{N}\sum_{i=0}^{N-1}{\pi}_{i}(t){\pi}_{i+r}(t) \rangle-\langle \frac{1}{N}\sum_{i=0}^{N-1}{\pi}_{i}(t) \rangle^2}{\langle \frac{1}{N} \sum_{i=0}^{N-1}{\pi}_{i}(t)\rangle-\langle \frac{1}{N}\sum_{i=0}^{N-1}{\pi}_{i}(t)\rangle^2}.
\end{equation}
The terms in the bracket are averaged over 100,000 independent trajectories. 
As the time scale of the maximal dynamic heterogeneity is comparable to the relaxation time of the system,
\cite{berthier2005numerical,whitelam2005renormalization}
  the time is fixed to be ${\tau}_{\text{pers}}$ of the system. 
In order to prove spatial aspects of dynamic heterogeneity in the crossover model, dynamic length scale is calculated (Fig.5(a)).
The dynamic correlation length, $\zeta$, which is defined as the distance that $C(r,{\tau}_{\text{pers}})$ falls to $1/e$, is chosen as the dynamic length scale.

\begin{figure}[t]
\begin{center}
\subfigure{
\includegraphics[angle=0,width=8.5cm]{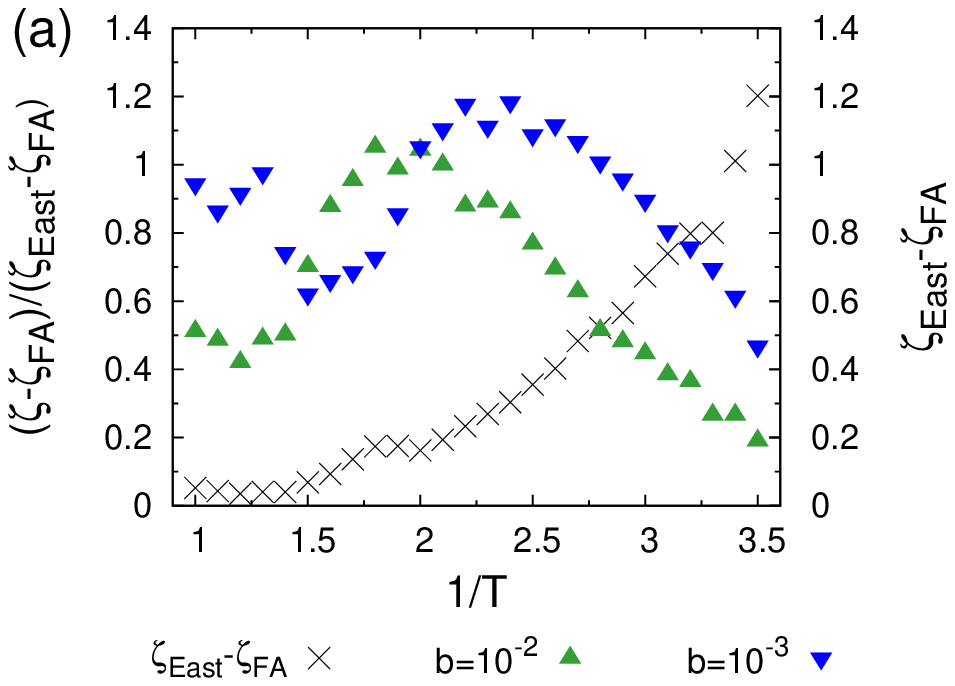}
}
\mbox{
\hspace{-37pt}
\subfigure{
\includegraphics[angle=0,width=8.8cm]{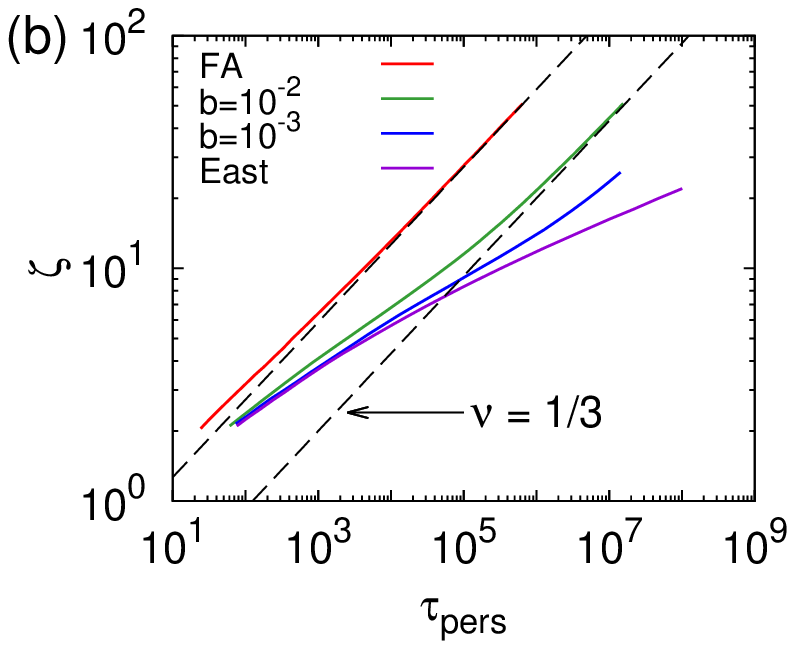}
}
}
\end{center}
 \vspace{-10pt}
 \caption{\label{fig6}
(a) Crossover of dynamic length scale from the FA model to the East model. 
(b) The relation between the dynamic correlation length and $\tau_{\text{pers}}$.
}
 
\end{figure}
Unlike  ${\tau}_{\text{pers}}$ or diffusion constant, the dynamic correlation length turns out to be a robust quantity that depends on the asymmetry parameter little (Fig.5(b)).
Nonetheless, there is a non-trivial difference between the two extreme cases - $\zeta_{\text{East}}$ is consistently larger than $\zeta_{\text{FA}}$ - and the intermediate models are in between.

In related studies,
\cite{berthier2005numerical,whitelam2005renormalization,kim2013multiple,klameth2014static}
  the dynamic length scale was defined as the scaling factor for the correlation function $C(r,t)$ or its Fourier transform $S(k,t)$. 
In Fig.5(a), it is demonstrated that our dynamic length scale can also act as a scaling factor and thus the two definitions are in agreement with each other. 
This also means that a different choice of the cutoff value for the correlation function would not affect our conclusion.

The crossover from the $b$ = 1/2 model behavior to the $b$ = 0 model behavior with respect to temperature is studied (Fig.6(a)). 
When the temperature is low enough $(T \le 0.35)$ $\zeta_{\text{East}}-\zeta_{\text{FA}}$ is much larger than statistical fluctuation, so smooth and approximately linear transition is obtained. 
The onset temperature of such crossover is similar to the crossover temperature obtained from ${\tau}_{\text{pers}}$ and diffusion constant. 

Intuition tells us that when temperature is sufficiently low, the active spins are sparse and each spin will be visited by a single active spin on average during the relaxation time, and therefore $\zeta \sim c^{-1}$. 
Combined with the FA models' scaling law $\tau_{\text{pers}} \sim c^{-3}$,
\cite{jung2004excitation} 
it is expected that the $b$ = 1/2 model will show scaling behavior of $\zeta \sim \tau_{\text{pers}}^{\nu}$ with $\nu = 1/3$ in low temperature, which is confirmed by our calculation as shown in Fig.6(b). 
The fragile-to-strong crossover is apparent in the intermediate models. 
The crossover temperatures would be those of ${\tau}_{\text{pers}}$, as the change in ${\tau}_{\text{pers}}$ over the temperature is much more significant than the change in the dynamic correlation length over the same range.

\section{Conclusion}
Evidence of dynamic heterogeneity in a theoretical model that shows the fragile-to-strong transition is investigated.
As the temperature decreases, functions related to dynamic heterogeneity make smooth fragile-to-strong crossover around a certain crossover temperature.
The dynamic properties such as the mean persistence time and the diffusion constant clearly show the crossover behavior.
The onset temperature of the transition is given as the function of asymmetry parameter, practically independent of the observable.
It monotonously decreases as the asymmetry of the kinetic constraint increases.
These observations are consistent with the entropic barrier mechanism proposed by Buhot et al.,
\cite{buhot2001crossover}
although our data shows that there is a room for improvement for better predictions around the crossover temperature. 

Using the relaxation time defined as the mean persistence time and the diffusion constant,
the fractional Stokes-Einstein relation is investigated.
One interesting feature is that there is no crossover behavior for the power law relations.
When the asymmetry parameter $b$ is fixed, the power law exponent is constant under the temperature change.
Above and beneath the transition temperature, the exponents are same in the range of error.
As expected, however, we find smooth transition of the exponents from 0.67 to 0.73 as the $b$ is reduced.
We note that there has been a recent debate on the nature of Stokes-Einstein breakdown in kinetically constrained models.\cite{blondel2014there,jung2013comment}

The dynamic length scale extracted from the spatial correlator $C(r,t)$ is also studied.
The correlation length experiences crossover from the value of the FA model to that of the East model.
Similar to the behavior in the triangular lattice gas (TLG) model system,\cite{pan2005decoupling}
the correlation length shows clear crossover with the relaxation time and recovers the power relation, $\zeta \sim \tau_{\text{pers}}^{1/3}$.
Note that, in our study, the dynamic length scale is much less sensitive to asymmetry of the kinetic constraint
than the transport properties such as the relaxation time and the diffusion constant.

The dynamic crossover has been explained as the consequence of the competition between entropy barrier of symmetric mechanism and energy barrier of asymmetric mechanism.
\cite{buhot2001crossover}
That the crossover temperatures of different physical quantities are the same function of asymmetry parameter seems to validate this explanation, 
although during the discussion on Eq.(4), this picture seems to be in need of a refinement around the crossover temperature where the symmetric process and asymmetric process occur in similar rate.

Finally, we would like to mention that fragile-to-strong crossover seen in our calculation has been experimentally observed in various glass-forming liquids and supercooled waters in confined state.\cite{ito1999thermodynamic,farone2004fragile,chen2006experimental,zhang2010fragile,mallamace2010transport}
The authors of these papers conjectured that there is a liquid-liquid phase transition occurring at the point of crossover.
Furthermore, the fragile-to-strong crossover behavior has its origin to local structure change such as change of hydrogen bonding environments. 
In our crossover model, however, the detailed information of local structure is coarse-grained out and the structure change is not considered as a critical origin of a crossover behavior.
In place of invoking local structure change, we use kinetic constraint and the asymmetric parameter to realize the crossover behavior.
Further scrutinies would seem necessary to reveal the microscopic mechanisms involved in the liquid-liquid transition.

\section*{Acknowledgments}
We would like to thank Sang-Won Park for useful discussions.
We also acknowledge the financial support from the Korean National Research Foundation (Grant Nos. NRF-2012R1A1A2042062 and NRF-2007-0056095) and
the Ministry of Education, Science and Technology, subjected to the project EDISON (EDucation-research Integration through Simulation On the Net, Grant No. 2012M3C1A6051724).
This work is also supported by the Brain Korea (BK) 21 Plus program. Computational resources have been provided by KISTI supercomputing center through Grand No. KSC-2013-C2-021.

\section*{Appendix: Derivation of Eq.(4)}
In our model system, we use the constraint that satisfies Eq.(\ref{condition_eq}).
The symmetric process can occur in both directions with the rate of $k_{+}\sim b$ and $k_{-}\sim (1-b)$, where $k_{+}$ and $k_{-}$ denote the local rate of flipping due to the excitation in positive direction and negative direction, respectively.
The total rate of symmetric process would be assumed in following way: ${k_{\text{total}}}^{-1}={k_{+}}^{-1}+{k_{-}}^{-1}=b^{-1}+(1-b)^{-1}=\frac{1}{b(1-b)}$.
Then, temperature dependent rate of symmetric process would be $\Gamma_{\text{sym}}\propto b(1-b)/{\tau}_{\text{sym}}$ where ${\tau}_{\text{sym}}$ is the temperature dependent relaxation time of symmetric model (FA model).
For the asymmetric process, we can rearrange Eq.(\ref{condition_eq}) to find the portion of asymmetric process: $P_i = (b(n_{i-1}+n_{i+1})+(1-2b)n_{i-1})\text{min}\left\{1,\text{exp}\left(-\frac{{\Delta}H}{T}\right)\right\}$.
In this way, we estimated that additional portion which comes from the asymmetric process would be proportional to $1-2b$.
The mean persistence time of the crossover model can be written in the following form:$
{{\tau}_{\text{pers}}}^{-1}=\Gamma_{\text{total}}=\Gamma_{\text{sym}}+\Gamma_{\text{asym}} \approx 4b(1-b){{\tau}_{\text{sym}}}^{-1}+(1-2b){{\tau}_{\text{asym}}}^{-1},\hspace{0.5cm}\left(0\le b\le 1/2\right)$.
The coefficient 4 in the numerator is adopted to reproduce the condition, ${\tau}_{\text{pers}}(b=1/2)={\tau}_{\text{sym}}$.

\bibliography{main}
\end{document}